\documentclass[twocolumn]{aastex61}
\hypersetup{linkcolor=red,citecolor=blue,filecolor=cyan,urlcolor=magenta}

\shorttitle{NGC 3516}
\shortauthors{}
\begin{document}
\title{Magnetic field in the Lobes of the Seyfert Galaxy NGC3516: Suggestions of a Helical field}

\author[0009-0000-1447-5419]{Salmoli Ghosh}
\correspondingauthor{Salmoli Ghosh}
\email{salmoli@ncra.tifr.res.in}
\affiliation{National Centre for Radio Astrophysics (NCRA) - Tata Institute of Fundamental Research (TIFR), S. P. Pune University Campus, Ganeshkhind, Pune 411007, Maharashtra, India}
\author[0000-0003-3203-1613]{Preeti Kharb}
\affiliation{National Centre for Radio Astrophysics (NCRA) - Tata Institute of Fundamental Research (TIFR), S. P. Pune University Campus, Ganeshkhind, Pune 411007, Maharashtra, India}
\author[0000-0000-0000-0000]{Esha Sajjanhar}
\affiliation{Department of Physics, Ashoka University, Rajiv Gandhi Education City, Rai, Sonepat, Haryana 131029, India}
\author[0000-0003-1933-4636]{Alice Pasetto}
\affiliation{Instituto de Radioastronom\'ia y Astrof\'isica (IRyA-UNAM), 3-72 (Xangari), 8701, Morelia, Mexico}
\author[0000-0001-8428-6525]{Biny Sebastian}
\affiliation{Space Telescope Science Institute, 3700 San Martin Drive, Baltimore, MD 21218, USA}

\begin{abstract}
We present polarization images from the Karl G. Jansky Very Large Array (VLA) and the Giant {Metrewave} Radio Telescope (GMRT) at 5.5, 10 GHz and 663 MHz of the changing-look (CL) AGN, NGC3516. A transverse gradient in the rotation measure (RM) is detected in the northern and southern kpc-scale lobes. Such gradients have typically been suggested to be signatures of a helical magnetic (B-) field. We detect circular polarization in the core and inner jet-knot of this source which is known to host a precessing radio jet interacting with emission-line gas. Soft X-ray emission from the \textit{Chandra X-ray Observatory} suggests the presence of a hot wind emerging from the nucleus of NGC3516. Taken together with the RM gradient, this presents a picture of jet+wind outflow in this Seyfert galaxy with the B-field confining both the jet and lobe emission. A magnetically driven outflow may in turn cause accretion disk warping and jet precession which is observed in the case of NGC3516. 
\end{abstract}
\keywords{Seyfert galaxies --- Polarimetry --- Radio interferometry --- Magnetic fields}

\section{Introduction}
A small fraction ($\sim$15\%) of active galactic nuclei (AGN) exhibit powerful, highly collimated jets on 100~kpc or Mpc scales, originating from the supermassive black hole–accretion disk system. These are classified as `radio-loud' (RL) AGN. In contrast, the majority ($\sim$85\%) of AGN are categorized as `radio-quiet' (RQ), defined by a ratio of radio flux density at 5~GHz to optical flux density in the B-band of $\leq$10 \citep{Kellermann1989}. Seyfert galaxies are RQ AGN characterized by bubble-like or lobe-like radio features typically confined within their host galaxies \citep[which are spirals or S0s;][]{Seyfert1943, Peterson1997}. 

Magnetic (B-) fields anchored to black hole-accretion-disk system are regarded as instrumental in jet production and collimation \citep{Blandford1977, Rees1978, Blandford1979}. Recent studies have found suggestions of the presence of helical magnetic fields in RL AGN, inferred from rotation measure (RM) gradients \citep{Asada2002, Gomez2008, Gabuzda2015, Pasetto2021}. Polarized emission undergoes Faraday rotation as it traverses the magneto-ionized media surrounding the source and intervening space. The intrinsic electric vector position angles (EVPA, $\chi_{int}$) are related to the observed values as $\chi_{obs}-\chi_{int}=\mathrm{RM} \lambda^2$, where $\lambda$ is the wavelength. RM is related to the electron density ($n_e$), line of sight B-field ($B_\parallel$) and the length of the Faraday rotating medium (L) by 
\begin{equation}
RM~(\mathrm{rad\, m^{-2}})=812\,n_e (\mathrm{cm^{-3}})\,B_\parallel\ (\mathrm{\mu G}) L\,(\mathrm{kpc})
\label{eq:RM}
\end{equation}

Sign changes in RM slices transverse to the jet or lobe can signify the toroidal component of a helical magnetic field \citep{Blandford2019, Wardle2018, Knuettel2017}. The absence of prominent jets in RQ AGN has posed challenges in predicting any organised B-field structures in their outflows. However, kpc-scale radio structures (KSRs) detected in several Seyfert galaxies \citep{Gallimore2006} offer a promising way for such an investigation. In this paper, we present multi-frequency polarization observations of the Seyfert galaxy with a KSR, viz., NGC3516, using data from the Karl G. Jansky Very Large Array (VLA) and the Giant {Metrewave} Radio Telescope (GMRT). NGC3516 is a nearby (z = 0.00883) barred spiral galaxy with its galactic disk inclined at an angle of 34$\degr$ \citep{Arribas1997, Ferruit1998}. Its radio-loudness parameter of 5.1 \citep{Sikora2007} places it firmly in the RQ AGN class. 

NGC3516 has been suggested to host a changing-look (CL) AGN \citep{Shapovalova2019,Mehdipour2022}. The source exhibited a type 1 spectrum until 2014, when the broad-line component significantly weakened and almost disappeared. A weak broad-line component began to re-emerge in 2017 and became sufficiently bright by 2020 \citep{Oknyansky2021}. This CL event is identified through variations in both the continuum and emission lines \citep{Ilic2020, Popovic2023}. The source is also associated with X-ray variability on timescales ranging from weeks to years, typically attributed to changes in the ionizing spectral energy distribution rather than obscuration \citep{Voit1987, Mehdipour2022}. During a 2019 flare in optical, X-ray, and UV wavelengths, high-ionization lines such as [Fe VII] and [Fe X] strengthened, and complex broad emission lines (H$\alpha$ and H$\beta$) emerged, indicating enhanced coronal activity and a reactivation of the AGN \citep{Shapovalova2019, Ilic2020}. NGC3516 is also known to exhibit a curved jet or lobe extending up to 4 kpc from the core \citep{Miyaji1992, Veilleux1993}. 

In this paper, the spectral index $\alpha$ is defined as S$_\nu\propto\nu^\alpha$, where S is the flux density and $\nu$ is the frequency. The adopted cosmology is H$_0=$67.8~km~s$^{-1}$~Mpc$^{-1}$, $\Omega_{mat}=$0.308, $\Omega_{vac}=$0.692.

\section{Radio Observations and Data Analysis}\label{sec:dataanalysis}
We acquired polarization-sensitive data at 5.5 GHz with the VLA in {BnA} array configuration 
on {5th February, 2018} with a resolution of $\sim 1\arcsec$ (corresponding to $0.192$ kpc at the distance of the source). {Details of all radio data are included in Table~\ref{tab:VLA-archival-obs}.} The {\tt Common Astronomical Software Application} \citep[CASA, version 6.6;][]{CASA2022} was used to perform the basic calibration steps, including initial flagging, instrument-specific calibration, initial delay, bandpass, and complex gains. We carried out additional flagging, and the final delay, bandpass, and gain calibrations. Finally, on the calibrated data, we performed the polarization calibration steps following the CASA-based polarization script for the VLA\footnote{\href{https://github.com/astrouser-salm/radio-imaging/blob/main/VLA_polarization_pipeline.py}{VLA\_polarization\_pipeline}}. For polarization leakage and angle calibration, the polarized calibrator 3C286 was used. The imaging was done using {\tt TCLEAN} in {\tt CASA} with {\tt Multi-term Multi-Frequency Synthesis} {\citep[MT-MFS;][]{Bhatnagar2013}} deconvolver and a {\tt Robust} weighting parameter of $+1.2$. After 3 phase-only and 2 amplitude+phase rounds of self-calibration, we created the Stokes I, Q, U and V images at a resolution of $1.4\arcsec \times 0.6\arcsec$ ($\sim$ 0.3 kpc $\times$ 0.1 kpc) at a position angle (PA) of $-74\degr$. The linear polarization fraction and the polarization angle values were given by $\sqrt{Q^2+U^2}/I$, and 0.5~tan$^{-1}$(U/Q). The circular polarization fraction values were obtained as $V/I$.

Polarization-sensitive data at 663~MHz were acquired using the upgraded GMRT (uGMRT) on 3 June, 2024. The calibration was carried out with the uGMRT polarization pipeline\footnote{\href{https://github.com/astrouser-salm/radio-imaging/blob/main/uGMRT_band4_polarization_pipeline.py}{uGMRT\_band4\_polarization\_pipeline}}. Leakage and polarization angle calibrations for the full-polar data have been carried out using the polarized calibrator 3C286. After 4 rounds of phase-only and 3 rounds of amplitude+phase self-calibration, we created the Stokes I, Q, U images at a resolution of $7\arcsec \times5\arcsec$ ($\sim$ 1.4 kpc $\times$ 1 kpc) at a PA of 22$\degr$, using a {\tt Robust} weighting parameter of $+0.5$.

\begin{figure*}
\centering
\includegraphics[width=8.9cm, trim= 130 20 110 20,clip]{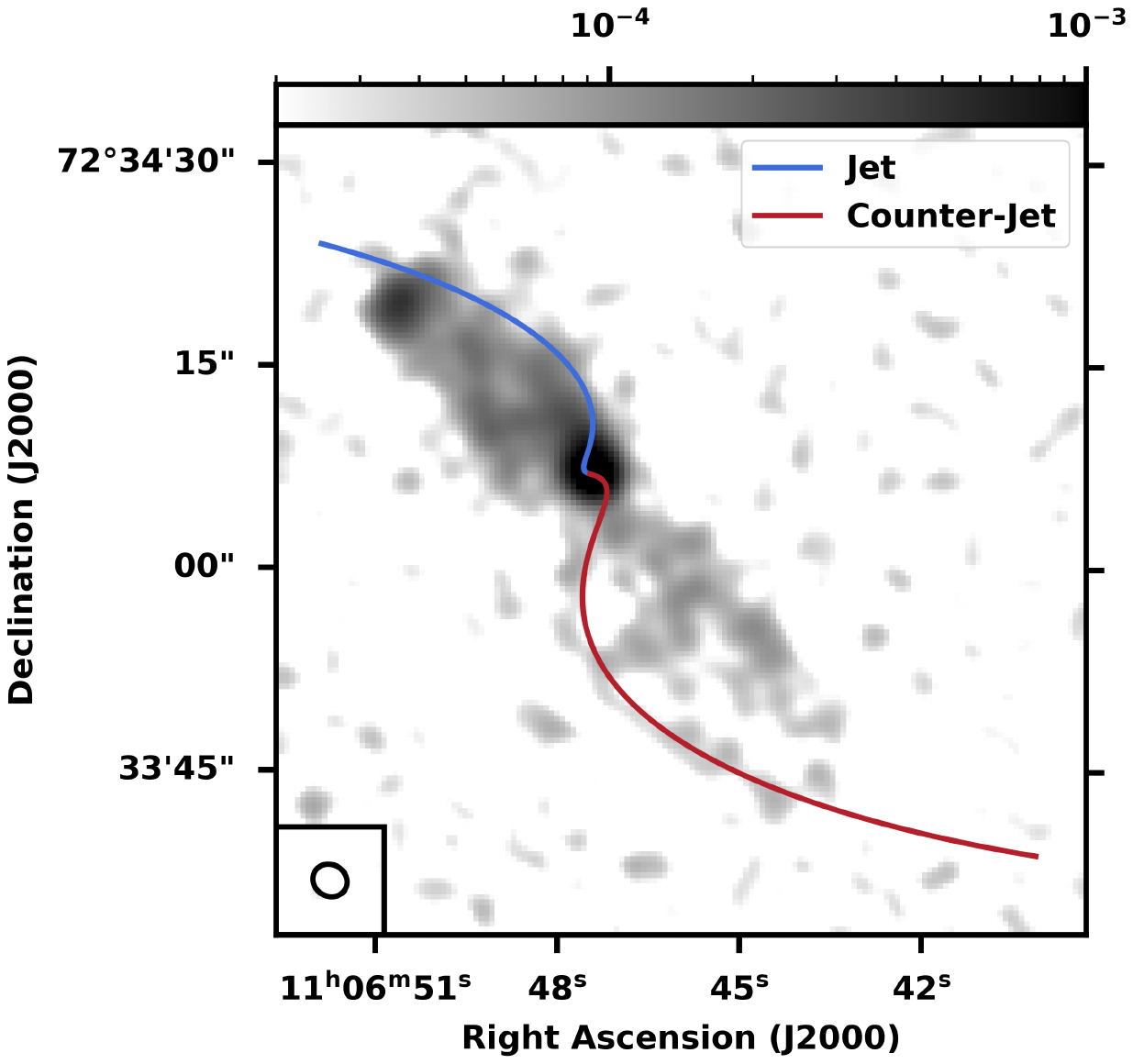}
\includegraphics[width=8.9cm, trim= 130 20 110 20,clip]{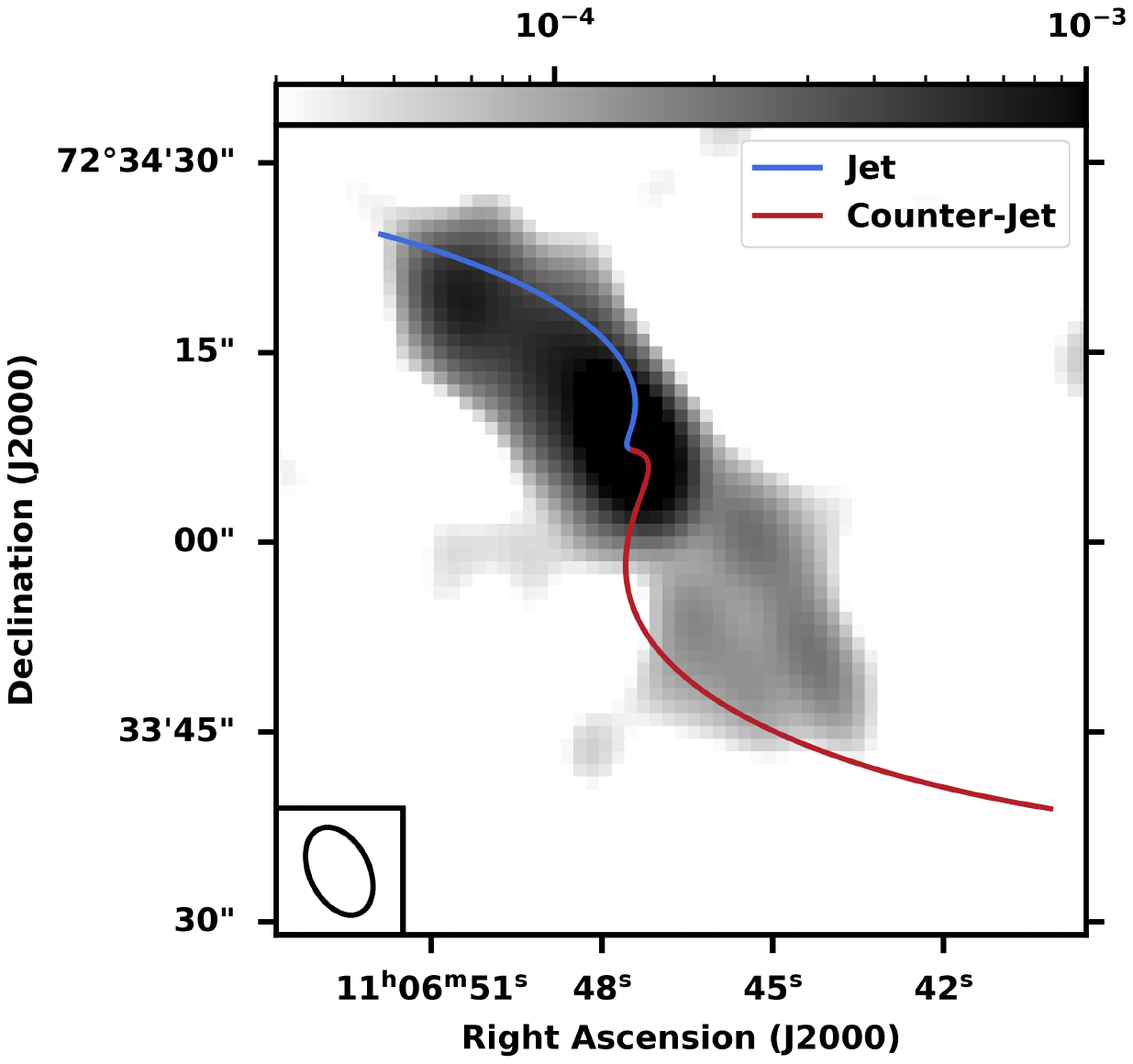}
\includegraphics[width=8.9cm, trim= 130 20 110 20,clip]{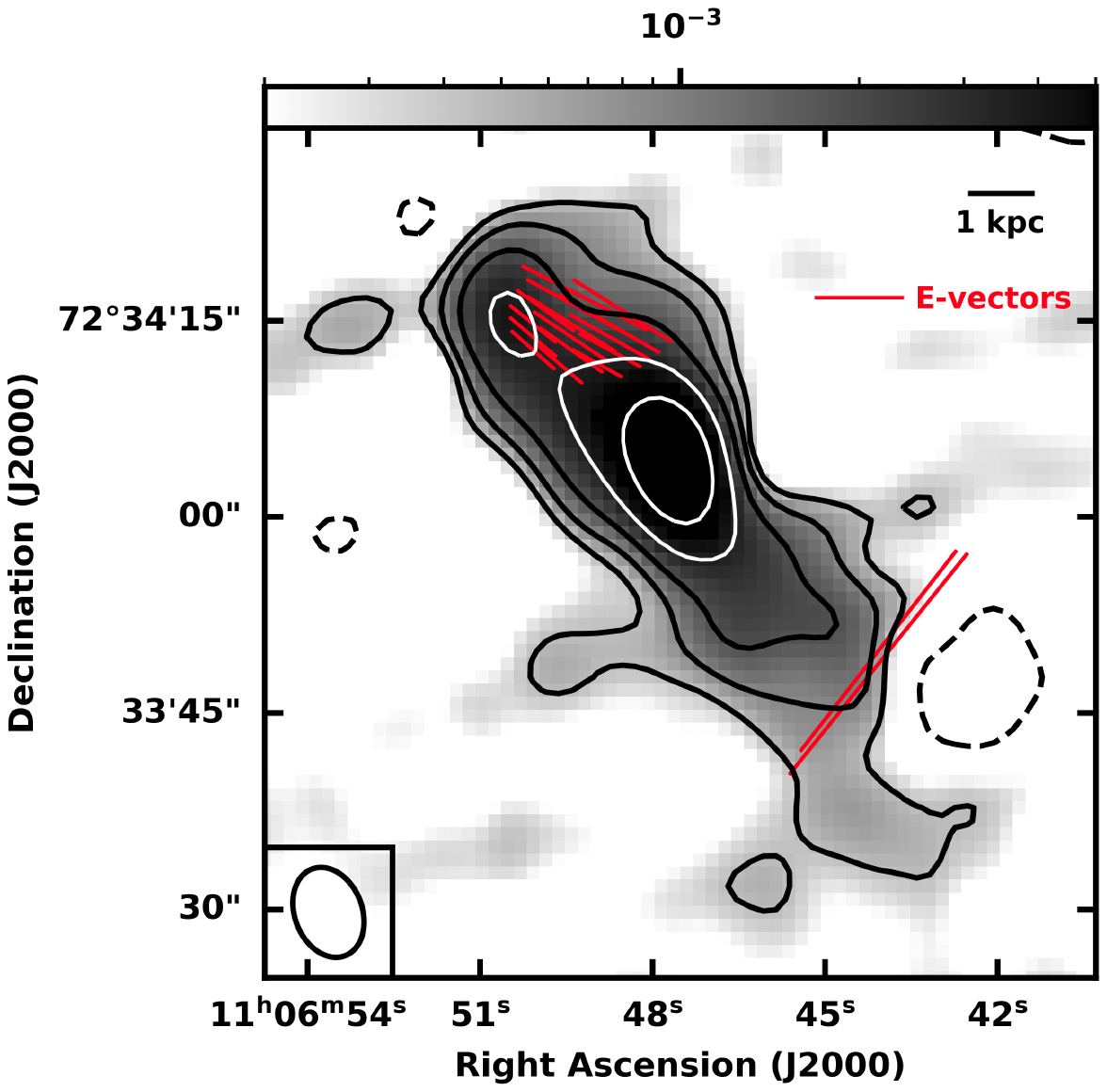}
\includegraphics[width=8.9cm]{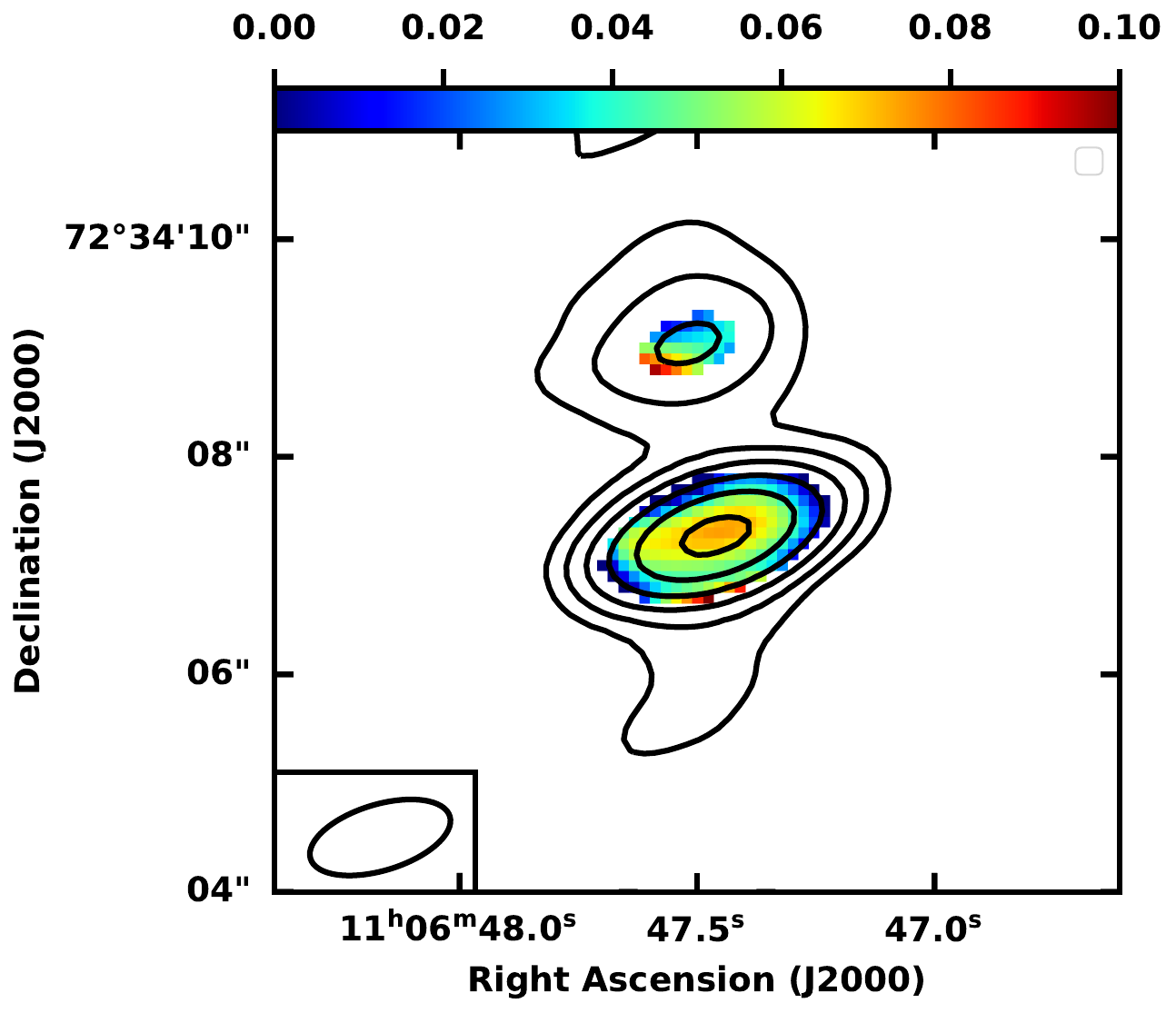}
\caption{\small Top left \& right: Best-fit precession model overlaid on 8.5 GHz and 10 GHz VLA images. Bottom left: The 663 MHz image of NGC3516 using uGMRT. The polarization vectors are shown in red ticks with 2.5$\arcsec$ proportional to a fractional polarization of 2\%. The contour levels are at 3$\sigma\times$ ($\pm1$, 2, 4, 8, 16) with $\sigma=150~\mu$Jy~beam$^{-1}$. Bottom right: Circular polarization from 5.5 GHz VLA image in colour with circular fractional polarization values ranging from $0-0.1$ or 0 to 10\%. The radio contours are at (2, 4, 8, 16, 32, 64)$\times 39~\mu$Jy~beam$^{-1}$ showing the core `A' and the first jet-knot `B'.}
\label{fig:fig1}
\end{figure*}

Archival VLA data at 1.5~GHz and 8.5 GHz with the A and C-array observed in January 1999 and April 2000, respectively, have also been reduced using the VLA pipeline {\tt{vlarun}} in {\tt{Astronomical Image Processing System}} \citep[{AIPS};][]{vanMoorsel1996} {to obtain total intensity images at these frequencies}. We performed a few rounds of both phase and phase+amplitude self-calibration to further reduce the r.m.s. noise in the images. 
{VLA 10 GHz D-array data observed on 9 November, 2019, whose polarization results are published as a part of a sample of 12 Seyfert galaxies in \citet{Ghosh2025}, have also been used in this paper. }

\begin{table*}
\centering
    \caption{VLA and GMRT observation details and observed parameters}
    \label{tab:VLA-archival-obs}
    \tabcolsep=0.1cm
    \begin{tabular}{cccccccc}
    \hline
    Date&Frequency&Array&Code&Beam Size \& PA&r.m.s. noise&Peak flux&Flux Density\\ 
    &MHz&&&&($\mu$Jy~beam$^{-1}$)&  (mJy~beam$^{-1}$)&mJy\\
    \hline
    20 Jan 1989&1500&A&AW221& 1.33$\arcsec$$\times$1.26$\arcsec$, $-85\degr$&38.2&4.55&$18.6\pm0.2$\\   
    4 Apr 2000&8460&C&AB942& 2.63$\arcsec$$\times$2.34$\arcsec$, $58\degr$&17.5&2.65&$9.1\pm0.2$\\
    5 Feb 2018&5490&BnA&17B-074&1.38$\arcsec$$\times$0.61$\arcsec$, $-74\degr$&13&2.94&$13.7\pm0.1$\\
    {9 Nov 2019}&{9998}&{D}&{19B-198}&{7.38$\arcsec$$\times$4.71$\arcsec$, $26\degr$}&{16}&{4.03}&{$8.1\pm0.4$}\\
    3 June 2024&663&GMRT&46\_086&7.12$\arcsec$$\times$5.26$\arcsec$, 22$\degr$&150&13.20&$50\pm2$\\
    \hline 
    \end{tabular}
\end{table*}

We constructed an RM image using the VLA 10~GHz data {\citep[from ][]{Ghosh2025}}, split into two sub-bands centred at 8.9 and 10.9 GHz, and the 5.5~GHz VLA data. 
We created all the images with an identical cell-size and image size, and convolved them with the poorest resolution beam, i.e., of $7\arcsec\times5\arcsec$ at PA$=22\degr$. The task `rmfit' in {\tt CASA} was used for obtaining the RM and RM error images. The final RM image presented here has been blanked with RM errors greater than 53~rad~m$^{-2}$. Additionally, we have subtracted the {Galactic} RM value of $-18$~rad~m$^{-2}$ \citep{Xu2014} along the line of sight to the source. We created RM slices perpendicular to the direction of the jet {(transverse slices)} and also along the largest RM gradient {(which is in the direction of the jet)}
in the northern lobe. {We have plotted both the continuous and discrete variations of the RM and its error along the slices. Each data point in the discrete measurement represents the median value of the RM sampled at 1/3 of the FWHM of the beam along the direction of the slice. To verify the presence of a gradient, we fit the discrete points along with their errors with a straight line having a positive or negative slope using the $\chi^2$ minimization technique. Due to large errors in the RM values in the transverse slices, we have rescaled the error values and computed the weighted reduced $\chi^2$.}

We have used archival data from the High Resolution Camera (HRC; Obs ID: 19519) as well as the Advanced CCD Imaging Spectrometer (ACIS; Obs ID: 2080) of the \textit{Chandra} X-ray Observatory {(see Section~\ref{dataavail}).} 
The {\tt Chandra Interactive Analysis of Observations} \citep[CIAO version 4.16;][]{CIAO} has been used for analysing the \textit{Chandra} data. The data were reduced in the standard manner by removing the bad detector pixels, taking the good time intervals, and running the $\tt{chandra\_repro}$ script. CIAO scripts were used for the conversion of photon counts to flux and separation of soft and hard X-rays in the ACIS image. 

\section{Results}
\subsection{Radio jet morphology and polarization}\label{sec:radiomorph}
The 8.5 and 10~GHz images with the VLA reveal the kpc-scale lobes with a distinct S-shaped symmetry (see Figure~\ref{fig:fig1}, top left \& right). We have identified the various jet and lobe components in the higher resolution VLA 5.5 GHz image (Figure~\ref{fig:fig2}, top left) as initially noted by \citet{Miyaji1992}. In particular, we note the core component `A', the first jet-knot `B' and the terminal region or hotspot `E'. The 663~MHz image from the GMRT reveals the presence of diffuse lobe emission of about 2 kpc beyond the VLA lobe in the southwest direction (see Figure~\ref{fig:fig1}, bottom left). Linear polarization is detected in the hotspot region as well as in the core in the VLA 5.5~GHz image with a fractional polarization of $20\pm7\%$ and $1.0\pm0.3\%$ and an EVPA of $57\degr\pm11\degr$ and $52\degr\pm13\degr$, respectively. These values are largely consistent with the fractional polarization limits for NGC3516 provided by \citet{Sebastian2020}.

The GMRT 663 MHz observations detect linear polarization in the northern and southern lobes (see Figure \ref{fig:fig1}, bottom left) with a fractional polarization of $5\pm1\%$ and $20\pm7\%$ at polarization angles of $56\degr\pm8\degr$ and $-37\degr\pm8\degr$, respectively. Linear polarization was also detected in both the lobes in our previous 10 GHz VLA D-array image \citep[see Figure~7 in][]{Ghosh2025}. 

{We used the VLA 1.5 (archival) and 5.5~GHz images to obtain the spectral index image (Figure~\ref{fig:fig2}, top right panel).} The radio spectral index image  reveals mostly optically thin emission {\citep[optical depth $<6$, $\alpha\leq+0.5$;][]{Cobb1993,Wardle2018}} in the source with {mean} $\alpha$ values of {$-0.28\pm0.01$} in the core (A), $-0.82\pm0.01$ in the jet-knot (B), {$-0.65\pm0.01$} in the northern hotspot (E), and $-0.75\pm0.06$ in the northern lobe {(A+B+C+D+E)}. Therefore, the inferred B-fields are assumed to be perpendicular to the polarization vectors \citep{Pacholczyk1970, Contopoulos2015}. We detect circular polarization (see Figure \ref{fig:fig1}, bottom right) in the core (A) with a fractional polarization of $6\pm1\%$. The northern jet-knot (B) also shows circular polarization with a fractional polarization of $5\pm4\%$, while linear polarization remains undetected here above the 3$\sigma$ threshold.    

\subsection{Precessing jet model} \label{sec:precessing-jet}
The S-shaped jet or lobe can be explained by invoking the precession of the jet ejection axis in NGC3516 \citep[e.g.,][]{Veilleux1993}. Such a precession could arise due to instabilities in the accretion disk \citep[e.g.,][]{Pringle1997}. We have used the precessing jet model from \citet{Hjellming1981} {to fit the VLA 8.5 GHz and 10 GHz images} therefore putting constraints on different properties of the jet like the opening angle of the precession cone, $\psi$ ($=30\pm15\degr$), the jet inclination to the observer's line of sight, $i$ ($=40\pm10\degr$), the instantaneous jet angle, $\theta$ ($=40\pm15\degr$), and the jet velocity with respect to the speed of light, $\beta$ ($=0.43\pm0.15$). {The best-fit model is based on a visual inspection of the best morphological match for the S-shaped structure in NGC3516.} Figure~\ref{fig:fig1}, top panel, shows the best-fitting precession model overlaid on the VLA 8.5 and 10 GHz images. 

\subsection{Constraints on jet velocity}
We derive the jet velocity using the jet-to-counter-jet surface brightness ratio ($R_J$), assuming the jet emission to be Doppler boosted and counterjet emission to be Doppler dimmed \citep{Laing-Bridle-2014}. The jet speed relative to the speed of light is {
\begin{equation}
    \beta = \frac{{R_J}^{1/(2+\alpha)} - 1}{\cos{\theta}({R_J}^{1/(2+\alpha)} + 1)}
\end{equation}}
where $\theta$ is the inclination angle of the jet. The northern lobe has an average spectral index of $\sim-0.8$ (see Section \ref{sec:radiomorph}). {We detect a jet-knot (B) 0.4 kpc north of the core, whereas no feature is detected at a similar distance to the south of the core.} Therefore using 3 times the r.m.s. noise of the 5.5 GHz VLA image about 0.4 kpc south of the core, as a limit to the counterjet brightness, and surface brightness of jet-knot B, a jet inclination angle of $30^{\circ}$ and an $\alpha$ of $-0.8$, we obtain a jet velocity of $0.48$c, consistent with the precessing jet model (Section~\ref{sec:precessing-jet}).  

\subsection{X-ray data}
With the \textit{Chandra}-HRC image, we observe an extension in the X-ray emission toward the north of the core (Figure~\ref{fig:fig2}, bottom left panel). The terminal point of this X-ray jet at $\sim300$ parsec from the core is behind the radio jet-knot at $\sim400$ parsec from the core. X-ray knots appear to be present at the edge of the north-eastern radio lobe. The longer-exposure ACIS-S soft X-ray (0.3 to 1 keV) image hints at a bowl-like wind emission in the X-rays (see Figure~\ref{fig:fig2}, bottom right panel). It also clearly reveals at least one bright X-ray knot along the edge of the north-eastern lobe. The nature of these X-ray knots needs to be probed with more sensitive X-ray data. Stacking all the ACIS-S data 
available for this source in the \textit{Chandra} archive, amounts to 386 ks of data but in High Energy Transmission Grating (HETG) mode. We fit the soft X-ray spectrum of the extranuclear `wind' region with a thermal and non-thermal component using Cash statistics. We obtained a good fit (reduced $\chi^2$ of 1.1) when we considered a range of column densities for the ionised winds ($10\pm5$, $23\pm15$ and $12\pm9$ $\times10^{20}$cm$^{-2}$), a power law component with photon index of $2.19\pm0.2$ and a blackbody with a temperature $2.3\pm0.7$ keV. This is consistent with the suggestion of multiple warm absorbers in NGC3516 \citep[e.g.,][]{Mathur1997, Mehdipour2010}.

\subsection{Equipartition B-field Estimate}\label{sec:equipartition}
Assuming the `minimum energy' condition, we have estimated the equipartition magnetic field ($B_\mathrm{min}$) as well as the electron lifetimes ($\tau$) in the radio lobes following the relations in \citet{OdeaOwen1987,vanderlaan1969}. We obtained the values of $C_{12}$ and $C_{13}$ from \citet{Pacholczyk1976} and varied the volume-filling factor ($\phi$) of the radio plasma in the lobes from 0.5 to 1 and the ion to electron energy ratio (k) from 1 to 50, in the equipartition equations. {From VLA 10 GHz D-array data,} we estimated the values of $B_\mathrm{min}$, $E_\mathrm{min}$ and $\tau$ to be ranging from $5-16\, \mu$G, $7-44 \times 10^{54}$ ergs, and $3.6-13.2$ Myr, respectively, in the northern lobe \citep[including the core and hotspot; {values for the core region have been presented in} ][]{Ghosh2025}. The relevant quantities are tabulated in Table \ref{tab:equipartition_results}. From the GMRT 663~MHz image, we estimate the in-band spectral indices of $-1.5\pm0.5$ and $-2.0\pm0.5$, and therefore an electron lifetime of $43\pm5$ Myr and $46\pm7$ Myr (considering $\phi$=k=1) in the primary radio lobe and the diffuse emission extending beyond (only detected at low-frequency observations with GMRT; see Figure~\ref{fig:fig1}, bottom left panel), respectively. While keeping the large errors in the electron lifetimes in mind, we can still attempt to derive an AGN duty cycle ($\epsilon\equiv t_{ON}/[t_{ON}+t_{OFF}]$) in NGC3516. We obtain $\epsilon\sim93^{+7}_{-24}$\%, which could suggest that this source is a `sputtering' AGN \citep[e.g.,][]{Sridhar2020}. 

\begin{table*}
\centering
\caption{Equipartition Estimates {using VLA 10 GHz data}} \label{tab:equipartition-results}
\begin{tabular}{cccccccc} \hline \hline
    Region&$\alpha$&$L_{\rm{rad}}$&$\phi$&k& $B_{\rm{min}}$&$E_{\rm{min}}$&$\tau$\\ 
    & (in-band)&(10$^{38}$~erg~s$^{-1}$)&& &(10$^{-6}$ G)&(10$^{54}$ ergs)&(10$^6$ yr)\\ 
    \hline
    Northern lobe (including core and hotspot)&-0.8&$5\pm5$&1&1&$5.4\pm1.5$&$7\pm4$&$13\pm3$\\
    &&&0.5&50&$16\pm5$&$33\pm18$&$3.6\pm1.4$\\
 
Total&-1.2&$12\pm1$&1&1&$6.9\pm0.3$&$17.2\pm1.4$&$10.6\pm0.5$\\
    &&&0.5&50&$21.3\pm1.0$&$81\pm7$&$2.6\pm0.2$\\
    \hline
    \end{tabular}

{\small Column 1: Source region for which energetics are estimated. Column 2: In-band spectral index value from VLA 10 GHz data. Column 3: Radio luminosity in units of 10$^{38}$~erg~s$^{-1}$. Column 4: Volume-filling factor of radio plasma. Column 5: Ion-electron energy ratio. Column 6: Equipartition magnetic field value in $\mu$G. Column 7: Energy at minimum pressure in units of 10$^{54}$ ergs. Column 8: Electron lifetime in Myr.}
\label{tab:equipartition_results}    
\end{table*}

\subsection{Kpc-scale Rotation Measure}\label{sec:RM}
{The RM image with errors less than 53 rad m$^{-2}$ is presented in Figure~\ref{fig:RMgradient}.} Interestingly, this image reveals a gradient both along the lobe and across the lobes. In the northern lobe, the RM values transition from positive to negative and back to positive, suggesting a reversal in the B-field direction from the lobe to the hotspot. A positive RM value indicates a B-field oriented {towards the observer}, while a negative RM corresponds to a B-field directed away from us. In the southern lobe, the RM values are predominantly negative but may also indicate a gradient.

To quantify these gradients, we extracted three slices: two transverse to the jet direction and one along the jet direction in the northern lobe. The RM gradients extend beyond {the FWHM of the angular resolution (7$\arcsec$ $\times$ 5$\arcsec$) along the direction of the slices}, attesting to their reality. The extension is closer to two beam-sizes in both the northern and southern lobes when the images are blanked with an RM error threshold of $\geq$60 rad m$^{-2}$; however, these images have not been presented here due to their larger RM uncertainties at the edges. 

Slice 1 {along the jet direction} in Figure \ref{fig:RMgradient} shows an RM gradient along the northern lobe {over a width of 16.5$\arcsec$ from north to south}. { The RM gradient is 2.4 times the major axis of the beam at a significance of 3$\sigma$. The statistical significance of the RM gradient has been calculated using the ratio of the magnitude of the largest difference in RM (at the two ends of the slices) to the largest error in RM along the slice \citep[see][]{Hovatta2012, Gabuzda2017}. However, the gradient along the jet} is likely to be affected by the presence of the hotspot and the ionised post-shock gas behind it, as noted by \citet{Miyaji1992}. Therefore, we do not discuss Slice 1 further. Slice 2 { which is transverse to the jet direction from east to west} in Figure \ref{fig:RMgradient} shows an RM gradient across the northern lobe {over a width of 8.5$\arcsec$ (which is 1.7 times the minor axis of the beam)} with values ranging from $-70$ to $+70$ rad~m$^{-2}$. {The significance of the gradient is 3.1$\sigma$.} Therefore, we consider the value of $\vert$RM$\vert$ as $35\pm35$ rad m$^{-2}$ for further calculations. Using a {screen length of 16$\arcsec$ ($\sim$3 kpc; the length of the part of northern lobe for which this RM is detected)} and the range of B-field mentioned in Section \ref{sec:equipartition} with the assumption that $B_{\rm{min}}=B_\parallel$ due to small jet inclination angle of 34$\degr$ ($B_{\rm{min}}/B_\parallel=\tan 34\degr \approx 0.7$, where $B_{\rm{min}}$ is the plane of sky B-field or $B_\perp$), we obtain $n_e < 0.005$ cm$^{-3}$ using Equation \ref{eq:RM}. This $n_e$ value is comparable to that of the ISM. 

{Slice 3 which is transverse to the jet direction from east to west in the southern lobe also shows a change in RM over a width of 7$\arcsec$ {(1.4 times the beam minor axis) at a significance of 3$\sigma$}. 
\citet{Hovatta2012, Livingston2025} have noted that an RM gradient of significance $\geq3\sigma$ spanning about $\geq1.4$ times the FWHM of the telescope beam can be considered significant, giving us confidence about the significance of our findings. Moreover, there have been similar findings of RM gradients for similar relative slice widths compared to the angular resolution, in AGN as well as protostellar jets and counterjets \citep[e.g.,][]{Gabuzda2017, Kharb2009, Kamenetzky2025}. Future higher resolution and more sensitive polarization images at multiple frequencies will be able to confirm these results with greater confidence.}

\section{Discussion}
\subsection{Helical magnetic field}\label{sec:helicalfield}
The inferred B-field is toroidal {(i.e., perpendicular to the jet direction)} in the $\sim 1\arcsec$ VLA core, considering the jet to be along the north-south direction. We detect {significant} circular polarization in the core `A' and {marginal circular polarization} in the jet-knot `B'. The circular polarization may arise due to Faraday conversion of the linearly polarized signal \citep{Beckert2002,Sullivan2013}, which is suggested to be due to the presence of a helical B-field \citep{Ensslin2003,Irwin2015,Irwin2018}. The transverse RM gradient in the kpc-scale radio lobes of NGC3516 supports the presence of a toroidal B-field component while a poloidal field component is observed in the VLA 10 GHz image of \citet{Ghosh2025}; taken together a kpc-scale helical B-field \citep[e.g.,][]{Gabuzda2015,Christodoulou2016,Pasetto2021} is implied in the lobes of NGC3516. Both the longitudinal (Slice 1 in Figure \ref{fig:RMgradient}) and transverse (Slice 2 in Figure \ref{fig:RMgradient}) gradients in the northern lobe display a transition from negative to positive RM values, with the magnitude of the RM remaining nearly constant. This consistency implies that the B-field strength and electron density are relatively uniform in the northern lobe, indicating only a change in B-field direction. A similar gradient is observed in the southern lobe with an opposite slope (Slice 3 in Figure \ref{fig:RMgradient}). {Hence, the two lobes show reversed gradients transverse to the direction of the jet.} This behavior is consistent with observations of other sources exhibiting helical magnetic fields, where the B-field structure manifests in opposite orientations along the approaching and receding jets/lobes \citep[e.g.,][]{Kamenetzky2025}. This implies that the helical B-fields are persistent in this source from parsec to kpc-scales, suggesting the outflow to be magnetically driven. 

Based on X-ray spectroscopic observations, \citet{Mehdipour2019} have suggested the presence of a jet+wind model in Seyfert galaxies with magnetic fields being poloidal in the jets and toroidal in the winds. The jet can be passing through the surrounding hot gas or a magnetized sheath acting as a wind \citep{Blandford2019}. The wind may act as a Faraday-rotating medium \citep[also see,][]{Kravchenko2020}. This suggestion is favoured by our radio polarization study of NGC3516. Moreover, as suggested by \citet{Lai2003}, a magnetically driven outflow can exert torques on the accretion disk, leading to warping and jet precession, which is indeed observed in NGC3516.

\subsection{Constraints on the depolarizing medium}
The medium depolarizing the radio source can be thermal/ionised gas of various geometry either present between the source and the observer or can be mixed with the source plasma, and the phenomena are thus called the `external' and `internal' depolarization, respectively.
Synchrotron photons moving through external media will give rise to a fractional polarization of 
\begin{equation}
p(\lambda^2)=p_i\,\exp{\{-2\,(812\,n_eB_\parallel)^2\,d\,R\,\lambda^4\}}
\label{eq:externaldepolarization}
\end{equation}

where `$p_i$' is the intrinsic fractional polarization, `$R$' is the {length along our line of sight} in kpc, `$d$' is the fluctuation-scale in the Faraday rotating medium in kpc and  $\lambda$ is wavelength in meters \citep{Burn1966,vanBreugel1984}. We analysed polarization data at 663 MHz, 5.5 GHz and 10 GHz, and the fractional polarization obtained for the northern lobe of NGC3516, after accounting for beam depolarization \citep[see][]{Contopoulos2015}, is $5.7\pm1.5\%$, $13\pm4\%$ and $17\pm4\%$, respectively. We tried solving for $d$, $R$ and $p_i$ by fixing the $n_e$ and $B$ values from our prior estimates. Considering an external Faraday screen of hot X-ray gas of $n_e\sim10^{-3}$ cm$^{-3}$ \citep[e.g.,][also see Section~\ref{sec:RM}]{Gizani2010, Ghosh2023} results in implausible values of the unknown parameters (e.g., $d>R$, $p_i<15\%$) with $B_\parallel = 5-15$~$\mu$G or $>17$~$\mu$G, in Equation~\ref{eq:externaldepolarization}. However, $B_\parallel = 16$ $\mu$G (which implies $\phi=0.5$ and $k=50$, see Table~\ref{tab:equipartition-results}) results in plausible values with $p_i$ = 17.3\%, $R$ = 21 kpc, and $d$ = 3 kpc. 

If the thermal plasma is mixed with the radio-emitting non-thermal plasma, the total B-field becomes $B_{\rm{tot}}^2=B_{\rm{uni}}^2+B_{\rm{rand}}^2$, where $B_{\rm{uni}}$ and $B_{\rm{rand}}$ are the uniform and random but isotropic B-field components. The complex fractional polarization is then given as 
\begin{equation}
P(\lambda^2)=p_i\frac{1-e^{-S}}{S}
\label{eqn3}
\end{equation}
where 
\begin{equation}
S=2(812\,n_e\,B_{\rm{rand}})^2\,d\,L\,\lambda^4-1624i\,n_e\,B_{\rm{uni}}\,L\,\lambda^2
\label{eqn4}
\end{equation}
and
\begin{equation}
p_i=p_i^\prime(B_{\rm{uni}}^2/B_{\rm{tot}}^2).
\end{equation}
$L$ is the Faraday depth here, and $p_i^\prime=\frac{3-3\alpha}{5-3\alpha}$ is the theoretical polarization fraction for synchrotron radiation in optically thin regions when $B_{\rm{rand}}=0$ \citep{Burn1966, Pacholczyk1970, Sullivan2013}. Considering $\alpha = -0.8$ for the northern lobe, $p_i =17\pm2$, we obtain $p_i^\prime\approx73\%$ and $B_{\rm{rand}}/B_{\rm{uni}} = 1.8\pm0.2$. Assuming, $B_{\rm{uni}} = \sqrt{3}B_\parallel$ \citep[see][]{Burn1966,Sullivan2013}, with $B_\parallel=16$ $\mu$G, we estimate $B_{\rm{uni}}=27.7$ $\mu$G, $B_{\rm{rand}}=49.7 \mu$G and therefore $B_{\rm{tot}}=57.3 \mu$G. Solving the imaginary part of Equations~\ref{eqn3} and \ref{eqn4} numerically, we obtained $n_e = 0.0017$ cm$^{-3}$, which is similar to the $n_e$ value obtained from the RM. This could be from a `mixing layer' in the outflow which would be consistent with the hot X-ray emitting gas in a wind and the outer layers of the lobe showing an RM gradient; i.e., the thermal plasma from the X-ray wind and the non-thermal plasma from the radio lobe could be mixed. 

The total mass of mixed gas present within the lobes can be estimated by $M_{tg} \sim n_e\phi m_H V$, where volume filling factor $\phi$ is assumed to be unity, total volume of the lobes  $V \approx (2.3\pm0.3)\times10^{66}$ cm$^3$ (assuming a cylindrical source volume), mass of ionised hydrogen $m_H = 8.4 \times 10^{-58}$ M$_\sun$. We obtain $M_{tg} \approx 3.2\times 10^6$ M$_\sun$. This is similar to the ionised gas mass obtained for other radio-quiet AGN \citep[e.g.,][]{Yoshida2002,Revalski2021,Silpa2022}.

\subsection{X-ray emission} 
\begin{figure*}
\centering
\includegraphics[width=8.95cm]{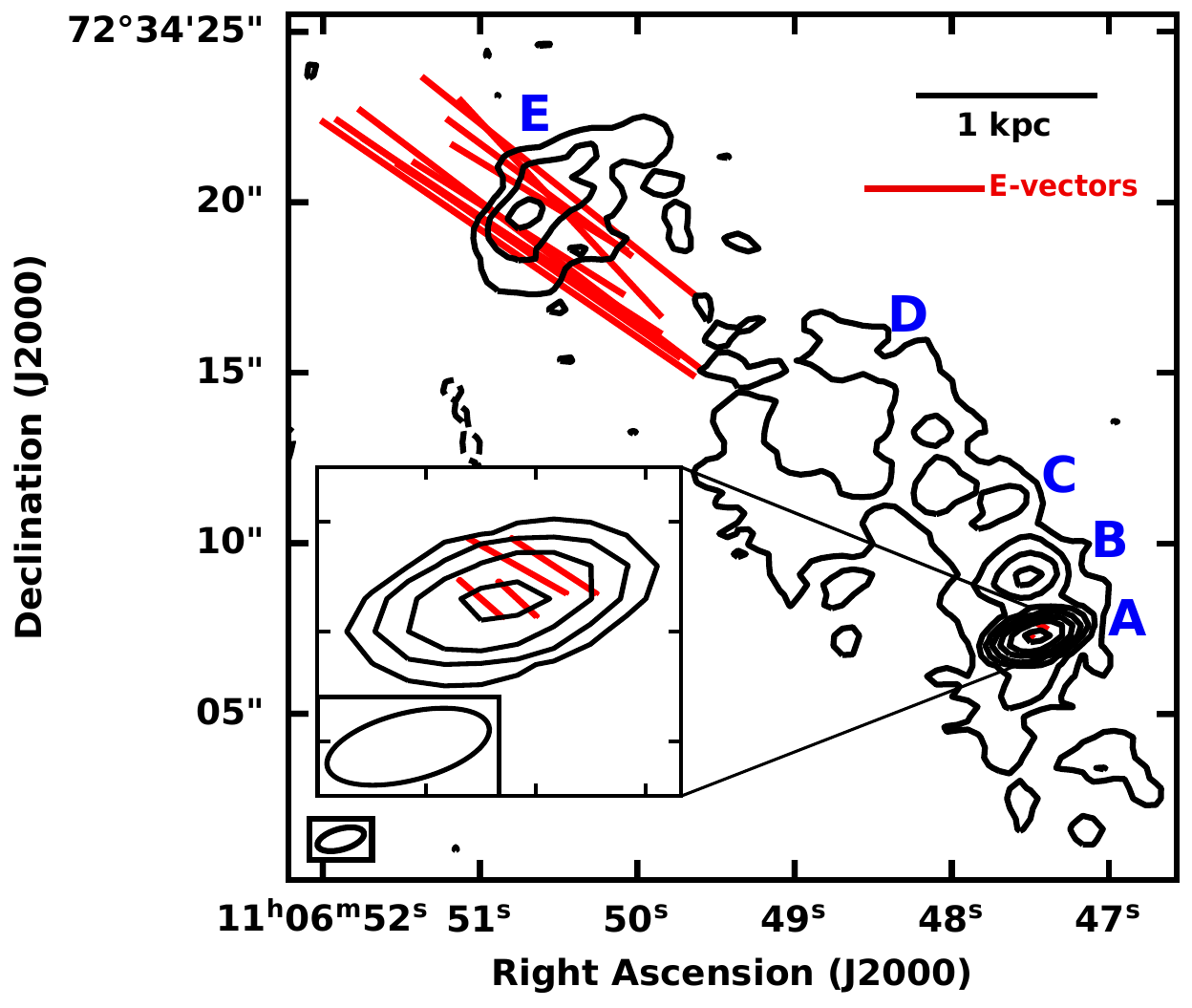}
\includegraphics[width=8.45cm]{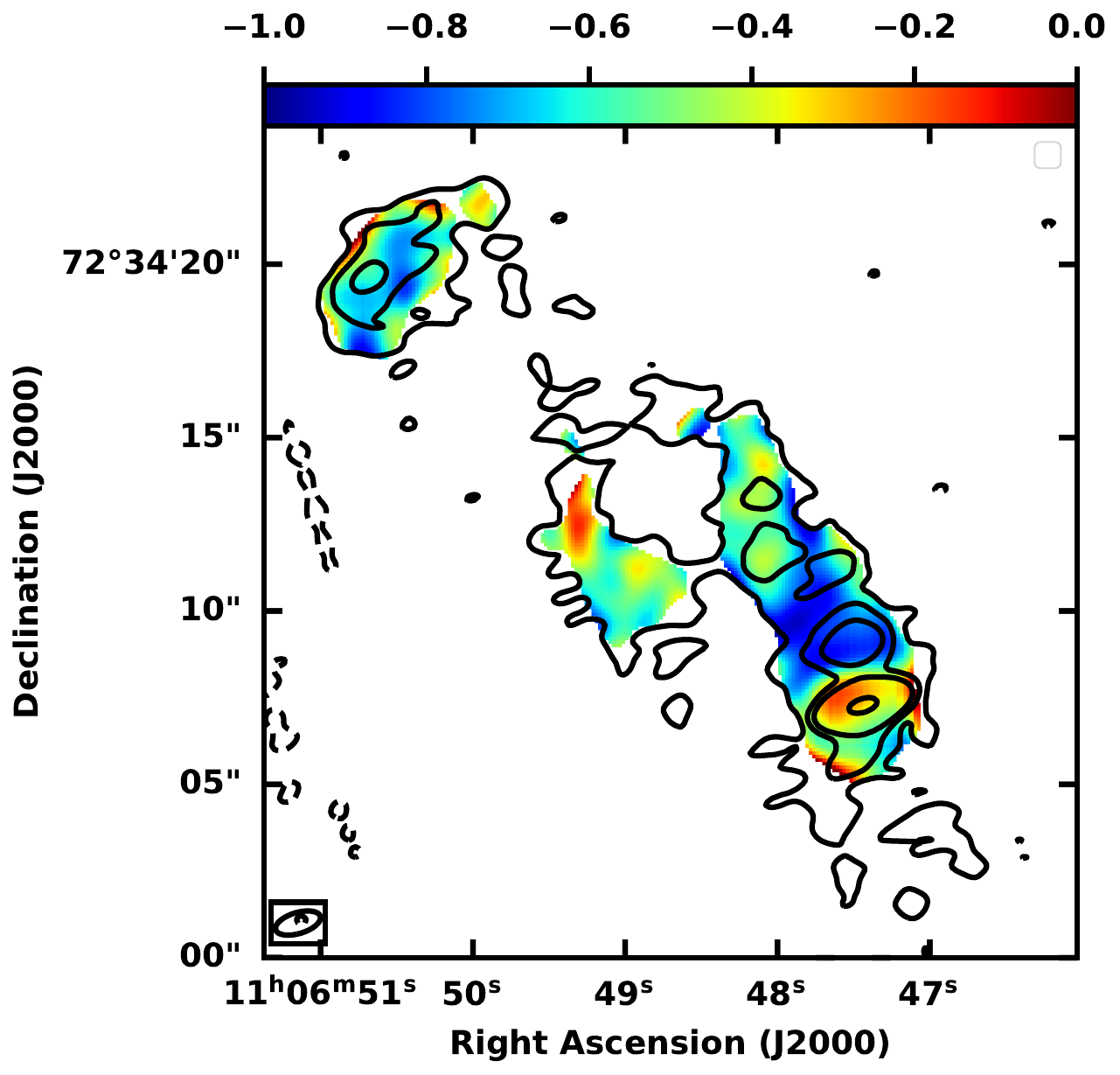}
\includegraphics[height=8.45cm]{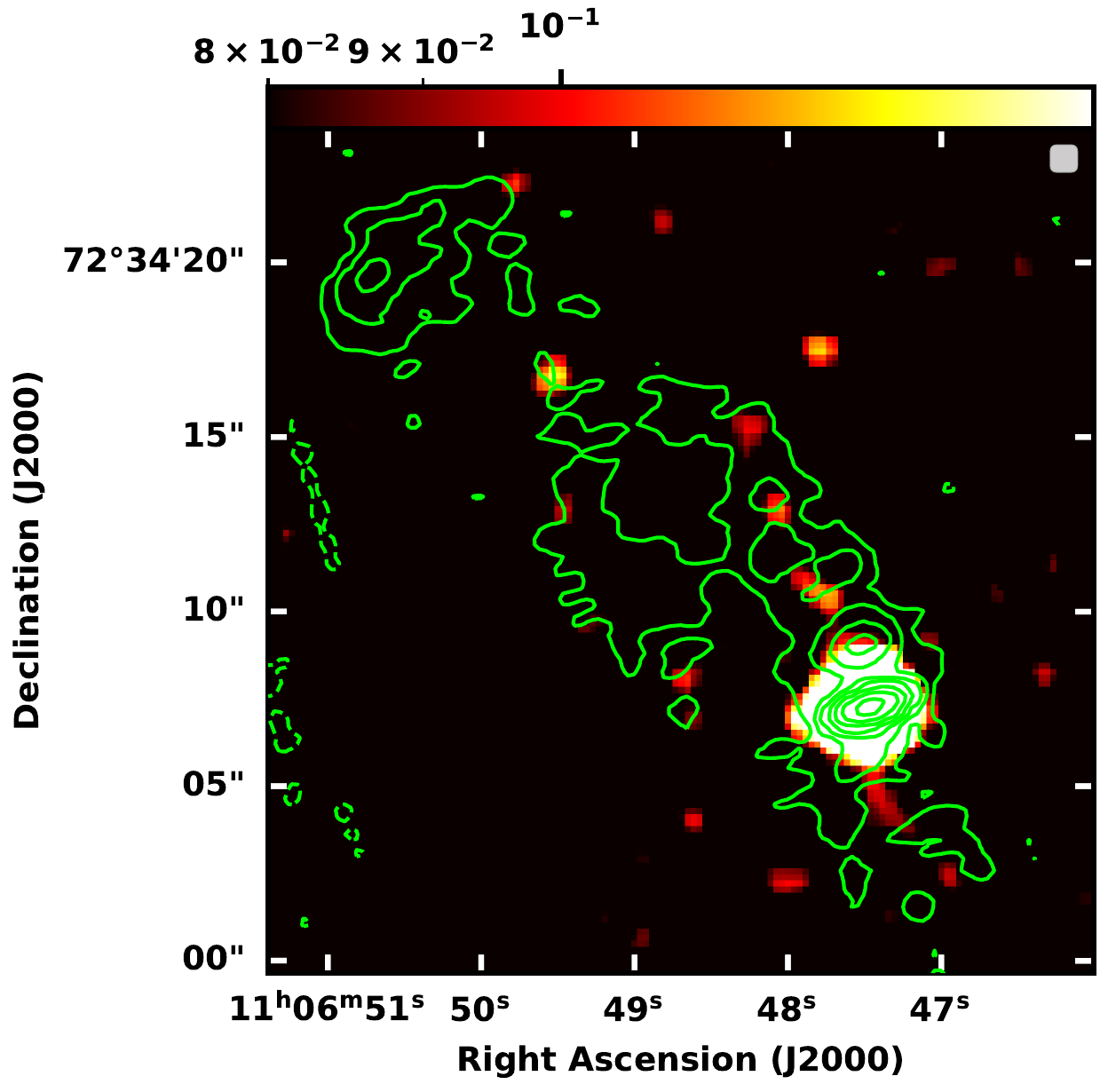}
\includegraphics[width=8.9cm, trim= 120 20 130 20,clip]{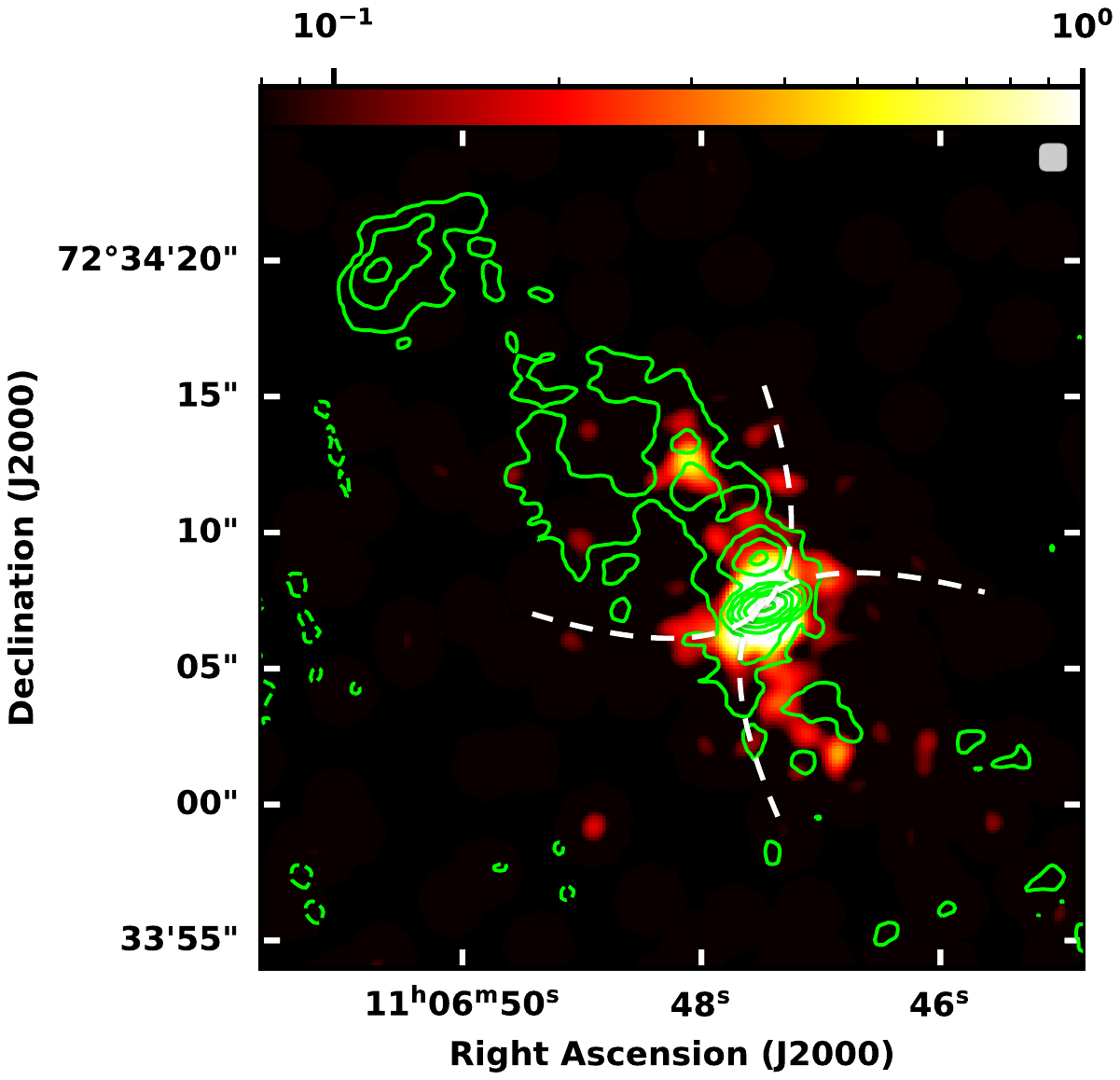}
\caption{\small Top left: 5.5 GHz VLA image of NGC3516. The contours are at level 3$\sigma\times$ ($\pm1$, 2, 4, 8, 16, 32, 64) with $\sigma=13 \mu$Jy beam$^{-1}$. The polarization vectors are shown in red ticks of length proportional to the fractional polarization of 5\% for every 2.5$\arcsec$. The different regions of the source are marked as A, B, C, D and E as noted by \citet{Miyaji1992}. The inset shows the EVPAs in the core of length of 5$\arcsec$ equivalent to the fractional polarization of 5\%. Top right: Spectral index image created using 1.5 GHz and 5.5 GHz data. The contour levels are at 3$\sigma\times$ ($\pm1$, 2, 4, 64) with $\sigma=13 \mu$Jy beam$^{-1}$. Spectral index values are in colour, ranging from $-1$ to 0. 
Bottom left: The $0.3-7$ keV \textit{Chandra} HRC X-ray image shown in colour in logarithmic scale of photon counts per sec. The radio contours are overlayed as the image on top.
Bottom right: The $0.3-1$ keV \textit{Chandra} ACIS soft X-ray image in colour with a total flux of 2.6 $\times$ 10$^{-12}$ erg cm$^{-2}$ s$^{-1}$. The radio contours in green are the same as the top. The white dashed line outlines the parabolic wind-like outflow.}
\label{fig:fig2}
\end{figure*}

\begin{figure*}
\centering
\includegraphics[width=9cm]{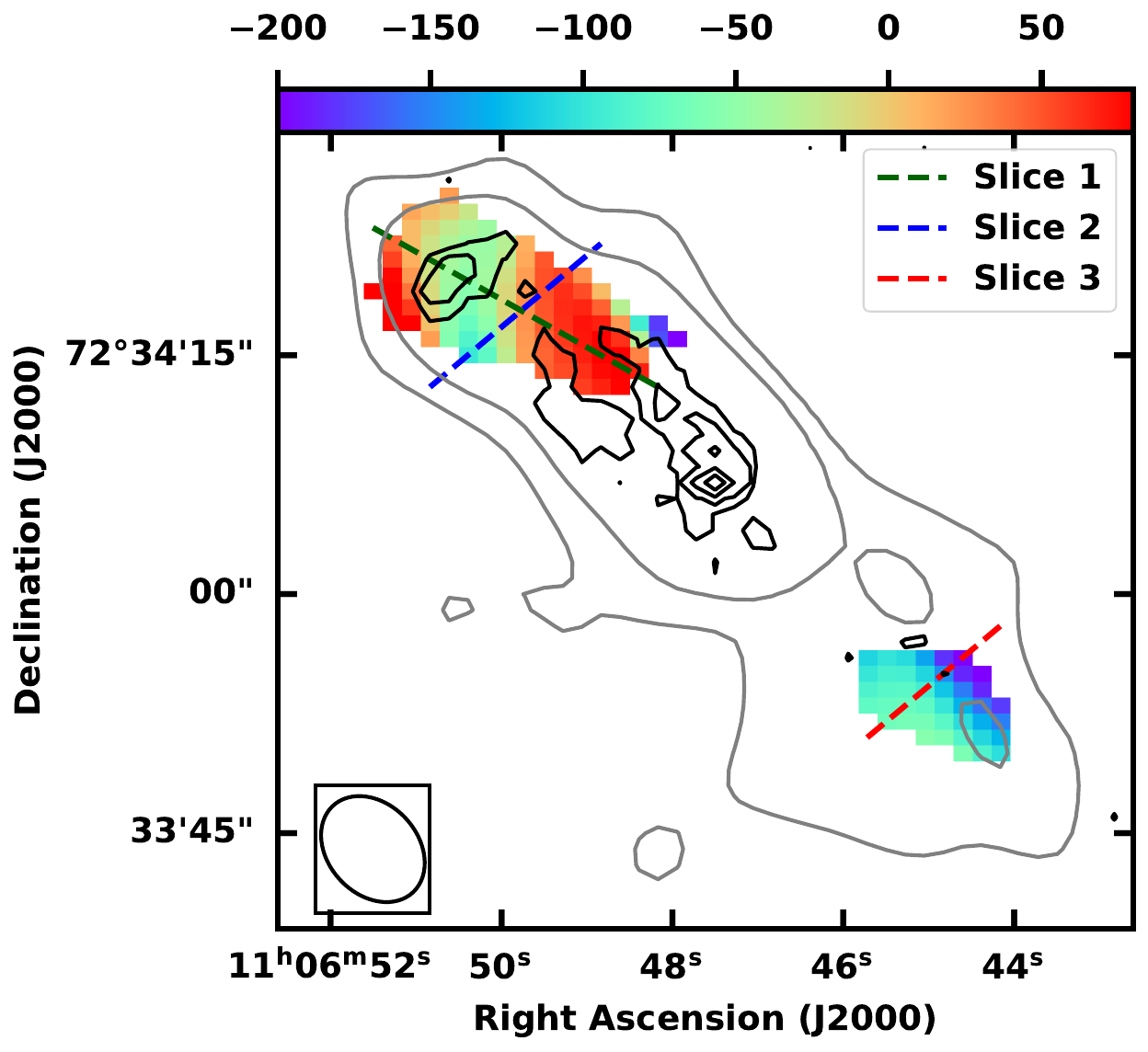}

\includegraphics[width=8cm]{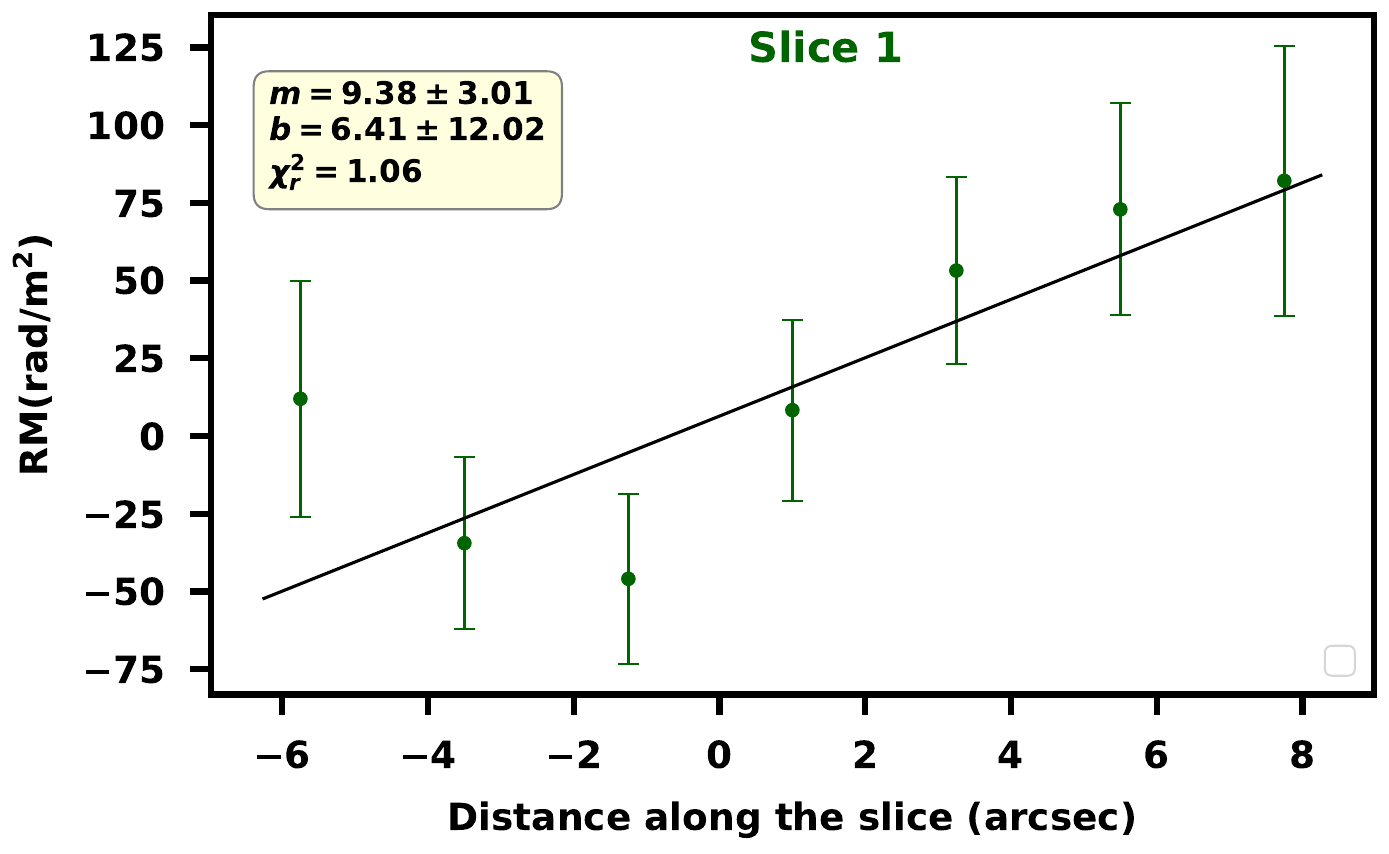}
\includegraphics[width=8cm]{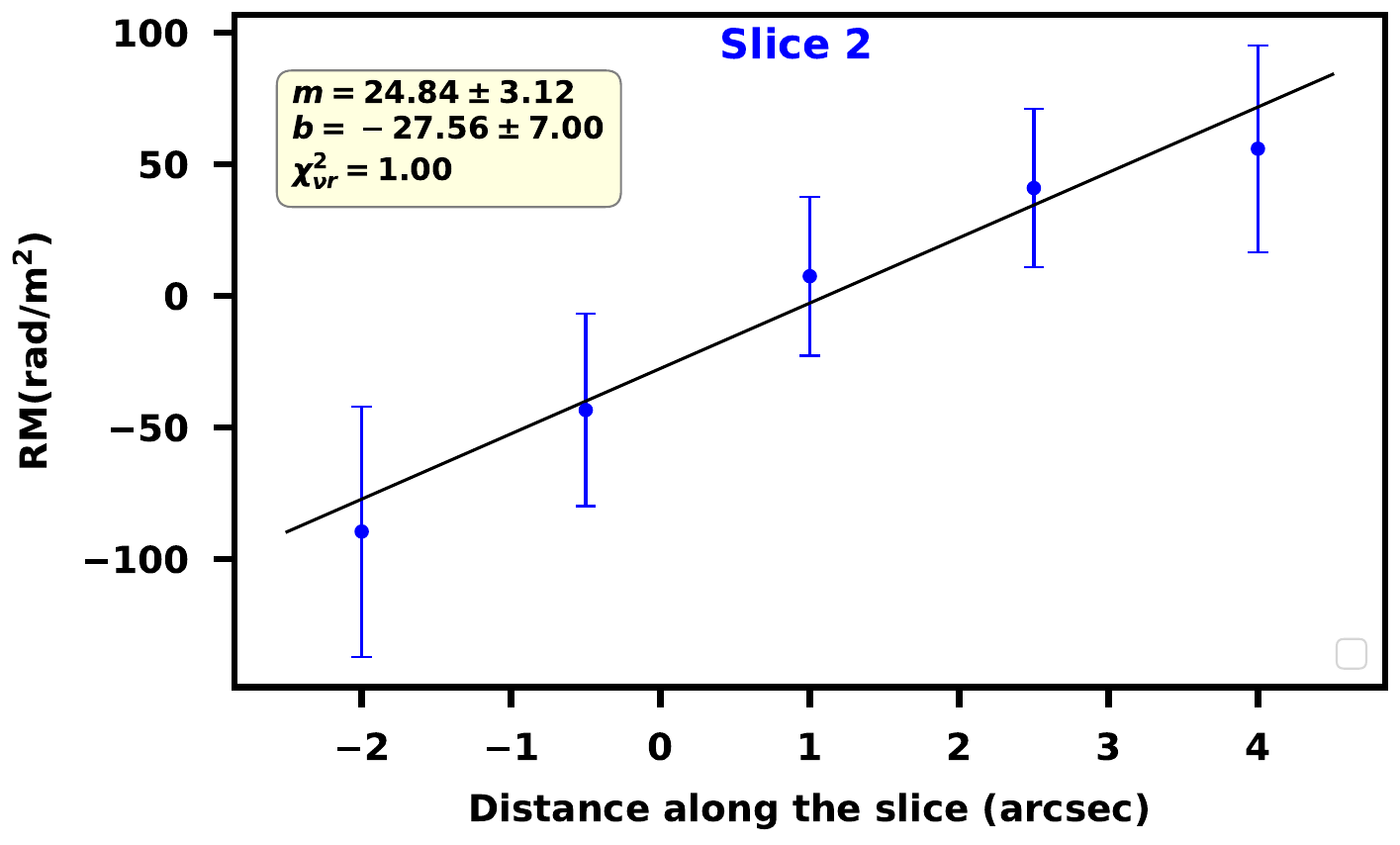}
\includegraphics[width=8cm]{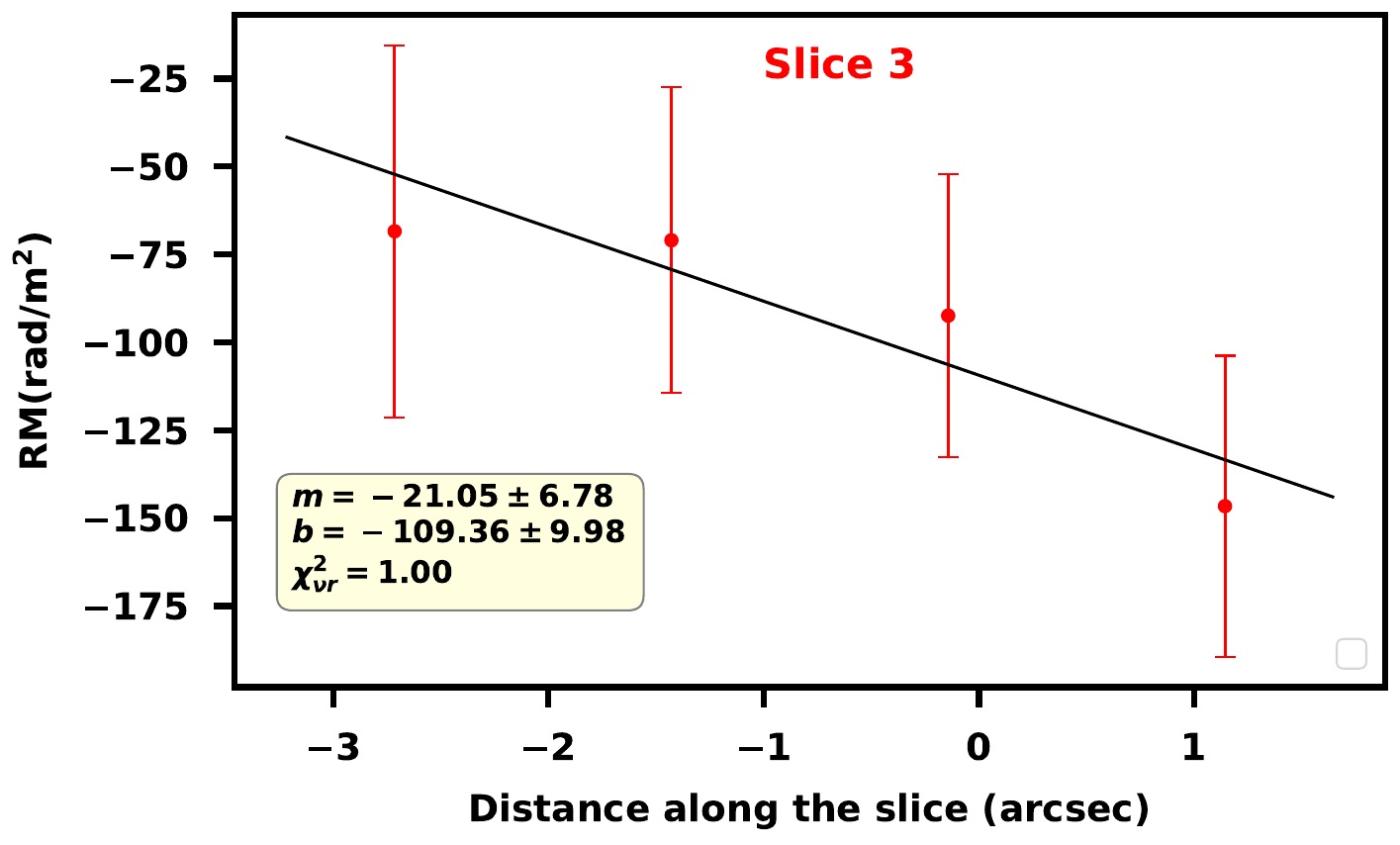}
\includegraphics[width=8cm]{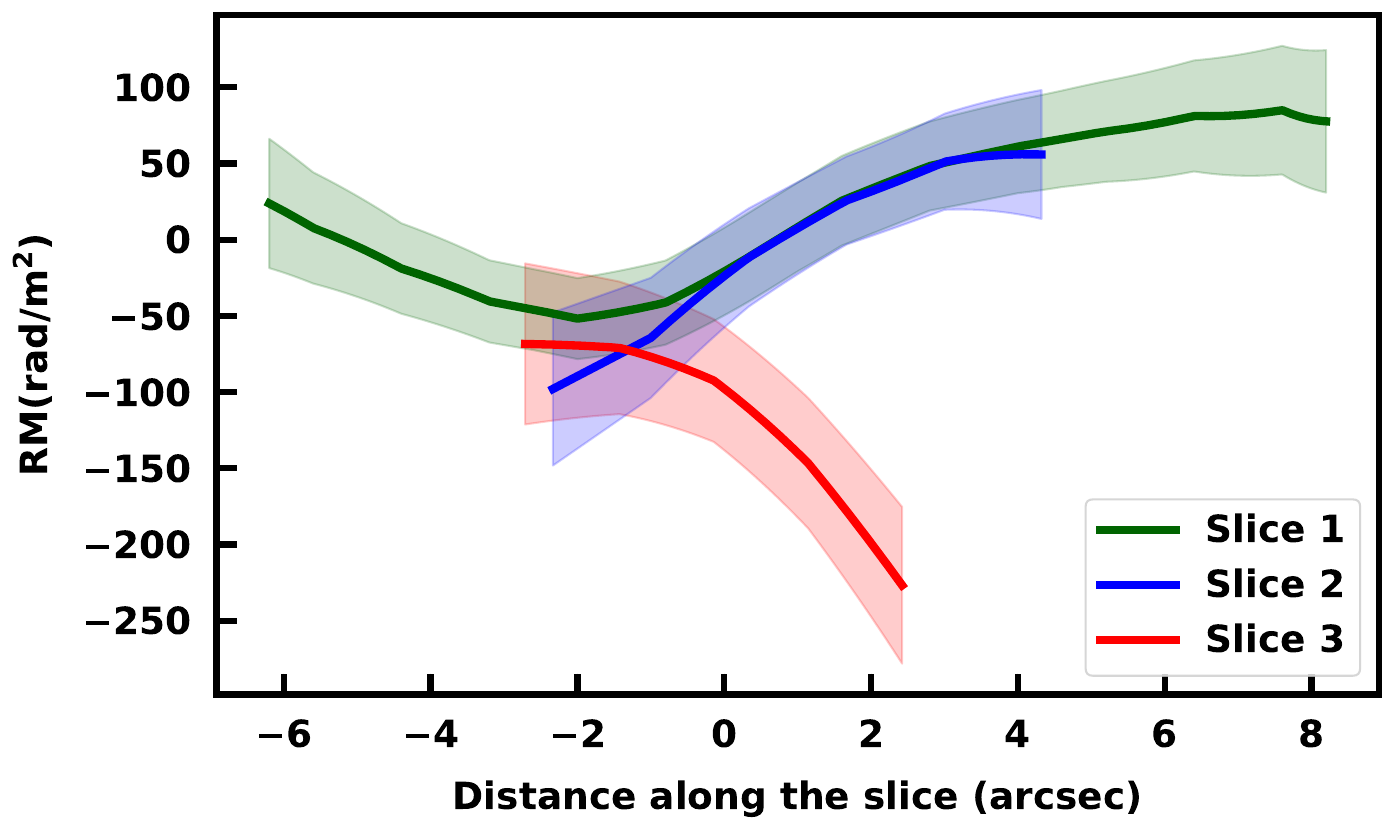}
\caption{\small Top: RM image shown in colour with values ranging from $-200$ to 80 rad/m$^2$. The VLA 10 GHz image contours are overplotted in grey at the levels 3$\sigma\times (1, 4)$ with $\sigma$ = 15.6 $\mu$Jy beam$^{-1}$. The VLA 5.5 GHz image contours are overplotted in black at the levels 3$\sigma\times (1, 2, 8, 32, 128)$ with $\sigma$ = 13 $\mu$Jy beam$^{-1}$. The slice axes are shown with dashed lines. {Middle left, middle right \& bottom left: Discretely sampled RM values with errors for slices 1, 2 and 3, respectively. The best-fit lines, shown in black, have been derived using reduced $\chi^2$ statistics. The slope (m), intercept (b) and the reduced $\chi^2$ ($\chi_r$) or weighted reduced $\chi^2$ ($\chi_{\nu r}$) have been noted individually in each panel}. Bottom right: Continuously sampled RM with errors shown as filled regions for the three slices.}
\label{fig:RMgradient}
\end{figure*}

The X-ray emission at around 0.3~kpc from the X-ray core, as seen in the \textit{Chandra}-HRC image, could be the site of bulk jet deceleration and particle re-acceleration \citep[e.g.,][]{Kharb2012}. The soft X-ray emission seen by \textit{Chandra} ACIS-S (Figure~\ref{fig:fig2}, bottom right panel) is reminiscent of the bowl-like wind emission observed in the Seyfert/starburst composite galaxy, NGC3079 \citep{Pietsch1998,Cecil2001}, and even the Milky Way \citep{Predehl2020}. Moreover, the coincidence of sub-kpc extended X-ray emission with radio lobes supports the idea of a magnetically-driven outflow \citep{Croston2008, Jones2020, Maksym2023}. Taken together with the radio data, the picture of a magnetically-threaded jet+wind structure emerges for the case of NGC3516. This jet+wind scenario, also consistent with our spectral model, was similarly inferred in the case of NGC3079 \citep{Cecil2001,Sebastian2019a} and the radio-loud AGN Cygnus A \citep{Ogle2025}.

\section{Summary \& Conclusions}
We summarise the primary findings of our polarization observations with the VLA and GMRT of the `changing look' Seyfert galaxy NGC3516 below.

\begin{enumerate}
\item The VLA and GMRT polarization images reveal the presence of transverse B-fields with respect to the outflow direction at the end of the north-eastern lobe, suggestive of B-field compression in the shocked terminal region. The inferred B-field in the core is perpendicular (i.e., a toroidal B-field component) to the VLA jet direction. Circular polarization is detected in the core as well as the inner jet-knot B at the $\sim1-10\%$ level. The spectral index image reveals steep spectrum ($\alpha=-0.75\pm0.06$) emission in the lobe and terminal region, with the radio core having a flatter ($\alpha=-0.28\pm0.01$) spectral index. 

\item Archival \textit{Chandra}-HRC data reveals $0.3-7$~keV X-ray emission from the core and inner jet region ($\sim$300 parsec from the core) suggestive of bulk jet deceleration and particle re-acceleration in an X-ray synchrotron jet. X-ray knots are also observed along the edge of the north-eastern lobe whose nature needs to be explored with more sensitive X-ray data. The soft ($0.3-1$~keV) X-ray emission observed by \textit{Chandra} ACIS-S exhibits a parabolic wind-like base around the radio lobes similar to what is observed in Seyfert galaxies like NGC3079. 

\item For the first time, an RM gradient is observed in the kpc-scale lobes of a Seyfert galaxy. The slope of the gradient in NGC3516 is reversed between the north-eastern and south-western lobes at a distance of $\sim$5~kpc from the core, strongly supportive of the presence of a large-scale helical magnetic field. This field could be threading the lobes in NGC3516. Taken along with the X-ray data, the picture of a jet+wind structure with poloidal fields in the jet and toroidal fields in the wind emerges, similar to that proposed by \citet{Mehdipour2019}. The detection of circular polarization in the core (which is the base of the jet and wind) at 5.5~GHz supports this finding. 

\item We estimate the electron lifetimes in the GMRT-detected lobe {at 663~MHz} to be $43\pm5$ Myr and in the diffuse radio emission extending beyond the south-western lobe to be $46\pm7$ Myr. This results in an AGN duty cycle of $\epsilon\sim93_{-24}^{+7}$\% in NGC3516, suggesting that this CLAGN could also be a `sputtering' AGN. 

\item Overall, NGC3516 appears to be a Seyfert galaxy with a jet+wind structure with a kpc-scale helical magnetic field threading the radio lobes. A magnetically driven outflow can cause accretion disk warping leading to precessing jets \citep[e.g.,][]{Lai2003}, as is observed in NGC3516.

\end{enumerate}

\section{Data Availability} \label{dataavail}
{All the VLA data are publicly available at \href{https://data.nrao.edu/portal/}{https://data.nrao.edu/portal/} under the respective project IDs mentioned in Table \ref{tab:VLA-archival-obs}. The proprietary period for the GMRT data is 18 months from the last date of observation. The GMRT data after that will be publicly available at \href{https://naps.ncra.tifr.res.in/goa/data/search}{https://naps.ncra.tifr.res.in/goa/data/search}. This paper employs a list of Chandra datasets, obtained by the Chandra X-ray Observatory, contained in the Chandra Data Collection (CDC) 390 ~\href{https://doi.org/10.25574/cdc.390}{(doi:10.25574/10.25574/cdc.390)}.}
.\section{Acknowledgements}
{We thank the referee for their suggestions, which have improved the manuscript significantly.} SG, PK acknowledge the support of the Department of Atomic Energy, Government of India, under the project 12-R\&D-TFR-5.02-0700. SG acknowledges the useful discussions with Ayushi Chhipa about X-ray data analysis. AP acknowledges support from UNAM DGAPA-PAPIIT grant  IA100425. This research has made use of data from the National Radio Astronomy Observatory (NRAO) facility. The NRAO is a facility of the National Science Foundation operated under cooperative agreement by Associated Universities, Inc. We thank the staff of the GMRT that made these observations possible. GMRT is run by the National Centre for Radio Astrophysics of the Tata Institute of Fundamental Research. This research has made use of data obtained from the Chandra Data Archive provided by the Chandra X-ray Center (CXC). {We have used Python modules, such as {\tt{NUMPY}}, {\tt{SCIPY}}, {\tt{UNCERTAINTIES}}, {\tt{ASTROPY}} and {\tt{MATPLOTLIB}} for producing the images and analysis.}

\bibliography{ms}{}
\bibliographystyle{aasjournal}

\end{document}